\documentclass[pre,aps,preprint,showpacs]{revtex4}
\usepackage{revsymb,latexsym,amssymb,amsmath,comment}
\usepackage{graphicx,psfrag,subfigure,stmaryrd}

\newcommand{\exval}[1]{\mbox{$\langle \, {#1}\, \rangle$}}
\newcommand{\be}{\begin{equation}}
\newcommand{\ee}{\end{equation}}
\newcommand{\bel}[1]{\begin{equation}\label{#1}}
\newcommand{\bea}{\begin{eqnarray}}
\newcommand{\eea}{\end{eqnarray}}
\newcommand{\ba}{\begin{array}}
\newcommand{\ea}{\end{array}}
\newcommand{\bef}{\begin{figure}}
\newcommand{\ef}{\end{figure}}
\newcommand{\e}{\mbox{e}}

\begin{document}

\author{Gunter M. Sch\"utz}
\affiliation{Forschungszentrum J\"ulich, IFF
D-52425 J\"ulich, Germany, and Interdisziplin\"ares Zentrum f\"ur Komplexe Systeme,
Universit\"at Bonn, Germany}
\email{g.schuetz@fz-juelich.de}
\author{Marian Brandau, Steffen Trimper}
\affiliation{Institute of Physics,
Martin-Luther-University, D-06099 Halle Germany}
\email{marian.brandau@physik.uni-halle.de}
\email{steffen.trimper@physik.uni-halle.de}
\title{Exact solution of a stochastic SIR model}
\date{\today }
\begin{abstract}

The susceptible-infectious-recovered (SIR) model describes the evolution of three species of individuals 
which are subject to an infection and recovery mechanism. A susceptible $S$ can become infectious with an
infection rate $\beta$ 
by an infectious $I$- type provided that both are in contact. The $I$- type may recover with a rate $\gamma$ and 
from then on stay immune. Due to the coupling between the different individuals, the model is 
nonlinear and out of equilibrium. We adopt a stochastic individual-based description where individuals are 
represented by nodes of a graph and contact is defined by the links of the graph. 
Mapping the underlying Master equation into a quantum formulation in terms of spin operators, the hierarchy of 
evolution equations can be solved  exactly for arbitrary initial conditions on a linear chain. In case of uncorrelated 
random initial  conditions the exact time evolution for all three individuals of the SIR model is given analytically. 
Depending on the initial conditions and reaction rates $\beta$ and $\gamma$, 
the $I$-population may increase initially before decaying to zero. Due to fluctuations, isolated regions 
of susceptible individuals evolve and unlike in the standard mean-field SIR model
one observes a finite stationary distribution of the $S$-type even for large population size.
The exact results for the ensemble averaged population size are compared with simulations for
single realizations of the process and also with standard mean field theory which is expected to be valid on large
fully-connected graphs. 

\pacs{05.70.Ln, 05.50.+q, 64.60.Ht, 75.10.Hk, 05.70.Fh}
\end{abstract}

\maketitle

\section{Introduction}

Infections produce further infections. This observation has long inspired
theoreticians to find simple tractable evolution equations to model such a situation.  
One traditional and rather simple approach 
is the so-called SIR model originally introduced in \cite{ker}, see also \cite{he} and \cite{mur}. 
Here a certain population is divided into three distinct classes: the susceptible, $S$, wherein the individual 
is healthy but is allowed to catch the disease; further there are the infectious, denoted as $I$ which is
infected and can transmit the disease and the 
recovered $R$ which is immune to further infection \cite{f1} 

Although the model is quite simple, it captures important features of the temporal dynamics of an infection cycle. 
In so far the model is appropriate to describe a well-localized disease outburst. Due to the coupling of the three 
different groups $S, I$ and $R$ the process is non-linear. Furthermore, as long as  there is an infectious population
the system is in an nonequilibrium state not characterized by any physical a priory principle such as
detailed balance. Despite its simplicity the SIR-model has not been solved 
exactly if fluctuations, which inevitably occur in a real system, are included in its description. In this paper we 
present such an exact solution using a mapping of the underlying Master equation into a quantum formulation. 
There appears a whole hierarchy of evolution equations for certain expectation values 
which can be closed and from which among other things the exact time evolution
of the expected population size for each class can be extracted analytically in closed form.

Our effort can be grouped in the permanent attraction exerted by modern biology and social science to 
understand the evolution of cooperative behavior. It is well known that in unstructured populations, 
natural selection favors defectors over cooperators. For that problem we also need the insight provided by 
mathematical tools. The SIR model offers a simple approach by a set of evolution equations \cite{mur,swrgc,
wgs,jl,mkcah,gmmr,smp}. To discuss the spreading of epidemics the SIR model can be implemented on a network 
\cite{z}, which is further discussed in \cite{aey,gk,v,ywbswz}. The general scheme and the properties of networks 
are elaborated in detail \cite{ab}. In such a network approach, individuals are represented by nodes
which are in either of the three states $S,I,R$. Contact between individual is modelled by links between the
nodes. For maximal connectivity, where each individual is in contact with every other, i.e. for the fully 
connected graph, one expects the deterministic standard mean field equations for the SIR model to be 
valid for large population size even if the infection and recovery are stochastic. In contrast, fluctuations 
are expected to cause substantial deviations from the mean field behaviour if  the connectivity is low. 
Hence it is highly desirable to study the opposite case of minimal connectivity as realized in a linear chain.
In the present paper we analyze the SIR model on  a linear chain based upon the master 
equation \cite{vk} which is reformulated in terms of second quantized operators \cite {doi,mg,gs,mts}. The method
enables us to apply the algebraic properties of spin operators to determine  
a closed set of evolution equations for higher-order cluster functions. These cluster functions describe 
groups of susceptibles which can be infected from 
the boundary of the region. The time evolution for the cluster functions can be closed which makes
the problem exactly solvable. This allows for a quantitative comparison with the mean-field solution 
as given by the deterministic standard SIR model and
also with the random behavior of single realizations of the process obtained from Monte-Carlo
simulation.

The paper is organized as follows.
In Sec. 2 we first define the standard deterministic SIR model and then introduce the stochastic dynamics 
that we consider to account for fluctuations.
In Sec. 3 we describe the mathematical apparatus required for obtaining the exact results.
This section can be skipped by readers not interested in the mathematical details. For an introduction
into the quantum approach used there we refer to the reviews \cite{mg,gs}. In Sec. 4 we present
the exact results for the expected population densities and compare
them with analytical results from the mean-field approach. In Sec. 5 we discuss
results of Monte-Carlo simulations for single realizations of the process . 
In Sec. 6 we finish with some conclusions.

\section{Stochastic SIR model on a graph}

Let us denote with $S(t),I(t)$ and $R(t)$ the number of susceptibles, infectious and 
recovered individuals. The total number $N$ is conserved
\begin{equation}
\label{con}
S(t) + I(t) + R(t) = N.
\end{equation}
In the traditional treatment of the SIR model the population strength is treated as a real
number and infection and recovery are governed by the nonlinear set of
coupled equations
\begin{eqnarray}
\frac{d S}{d t} &=& - \beta S I \nonumber\\
\frac{d I}{d t} &=& \beta\, S\,I - \gamma I \nonumber\\
\frac{d R}{d t}  &=& \gamma I.
\label{mfa}
\end{eqnarray}
The first equation describes the decrease of the susceptible population
through the infection of a susceptible individual by an infectious one. 
The loss is proportional to the infection rate $\beta$ and since by definition
$S$ and $I$ are non-negative, the loss is monotone. The second equation
describes the gain of the infectious population of individuals by infection of susceptibles
as described in the first equation and the spontaneous recovery with rate $\gamma$.
The last equation follows simply from the conservation of the total number of individuals.

These equations describe a deterministic evolution for each
population class which entirely neglects fluctuations and which offers no description
of the state of an individual member of the entire population. These equations may be
regarded as a mean-field treatment of some fluctuating random process and
therefore we shall refer to this standard SIR model as mean field SIR model.
In view of our further approach it is appropriate to introduce the 
population densities 
$n_X(t) = X(t)/N$ where $X$ stands for one of three classes $S,I,R$.
Obviously the densities satisfy $n_S(t) + n_I(t) + n_R(t) = 1$.

We now define a stochastic SIR dynamics that describes the state of each
individual. This description allows for
randomness and hence fluctuations in both the infection and recovery process.
In our individual-based version of the model the individuals are represented by the nodes of a graph.
For each node $i$ we introduce
state variables which specify the state of the node. For reasons that become clear below
we represent these state variables as occupation numbers which take value 0 or 1 as follows:
If node $i$ is in the susceptible state at time $t$, we say that $n_S(i,t)=1$. If node $i$ is in the infectious 
or recovered state then $n_S(i,t)=0$. Likewise we define occupation numbers $n_I(i,t), n_R(i,t)$
which by definition are subject to the constraint $n_S(i,t) + n_I(i,t) + n_R(i,t) = 1$.
With this definition we define the (random) population sizes of class of individuals
\bel{pop}
\hat{X}(t) = \sum_i n_X(i,t)
\ee
where the sum is taken over all nodes of the graph. Considering $N$ nodes
ensures a strict conservation law $\hat{S} + \hat{I} + \hat{R} =N$ analogous to
\eqref{con} for the deterministic SIR model.
Contact between two individuals is represented by
a link between two nodes. This defines the graph.

The stochastic dynamics of the model is realized by the following Markov process. 
A susceptible individual at a node $i$ becomes infectious after an exponentially distributed random time 
with rate $\beta I(i,t)$ where  
\bel{neighbor}
I(i,t) = \sum_{j( i)} n_I(j,t)
\ee
is the total number of infectious individuals $j(i)$ which are in contact with $i$ at time $t$.
This quantity is an integer random variable that depends on the current state of the system. 
On the other hand, an infectious individual at node $i$
recovers after an exponentially distributed random time with fixed rate $\gamma$. Once
an individual is recovered it remains so. All infection and recovery processes occur
independently of each other. 

Thus this stochastic process is in double contrast to the evolution studied in the
mean-field approach. There recovery and infection are deterministic and the
infection rate is
proportional to the size of the {\it full} population of infectious individuals.
The latter property is recovered in our individual-based approach if each
individual is in contact with every other,  i.e., if the underlying network
is the complete graph of $N$ nodes. If then in addition the population size
$N$ is send to infinity, one expects fluctuations to disappear by the law of 
large numbers. Hence the traditional SIR model as described by (\ref{mfa})
may be regarded as a deterministic limit of the evolution of our stochastic 
process on a complete graph in the thermodynamic limit of infinite
population size. 

In our stochastic model the main quantity of interest is the {\it expected} state
$\exval{n_X(i,t)}$ of a node $i$ at time $t$, given some initial distribution.
We shall focus on uncorrelated random initial distributions with some
given mean population size for infectious and susceptible individuals.
In this case the expectation value is independent of the node $i$ and we write
it in slight abuse of notation $n_X(t) = \exval{n_X(i,t)}$ and $X(t) =
\exval{\hat{X}(t)}$ \cite{f2}.
Moreover, in order to quantify and highlight the possible effect of 
fluctuations due to incomplete connectivity between individuals
we study the most ``non-mean-field'' setting possible.
I.e. we consider the lowest possible connectivity between individuals
which is realized in a periodic chain of $N$ nodes.
The dynamics of the model is then realized by the following transitions
on neighboring nodes
\begin{eqnarray}
I\,S &\to& I\,I \quad\quad \mbox{with rate}\quad \beta \nonumber\\
S\,I &\to& I\,I \quad\quad \mbox{with rate}\quad \beta \nonumber\\
I &\to& R \quad\quad\,\, \mbox{with rate}\quad \gamma\,.
\label{mod}
\end{eqnarray}
The first two processes mean that a single susceptible $S$ can catch the disease when it is in contact with
infectious individuals situated on the neighboring nodes. The last process in Eq.~\eqref{mod} characterizes the recovering 
process, where an infectious individual recovers and becomes immunized, independently of the
state of other individuals.

\section{Quantum Approach to Nonequilibrium Systems}

\subsection{Master equation in a quantum Hamiltonian representation}

Since the combined influence of noise and spatial degrees of freedom 
is an important issue in a theoretical understanding of biological and ecological processes \cite{rmf} 
the dynamics due to Eq.~\eqref{mod} is formulated in a master equation for the full
probability distribution of the process.
Here we use a very transparent 
method, the transformation of the master equation into a quantum language. This exact mapping enables us to 
get an exact solution for the process defined above. Since the method borrows techniques
from condensed matter and particle physics, we use well-established jargon that is slightly
different from that above. In particular, we shall refer to nodes of the graph as sites on a
lattice, and to the state variable $n_X(i)$ as occupation numbers by particles of type $X$.

Let us summarize briefly the most important steps, for a detailed
account of the approach see \cite{mg,gs}. The master equation for the joint probability 
$P(\vec n, t)$ reads
\begin{equation}
\partial_t P(\vec n,t)= \mathcal{L}P(\vec n, t)\,. 
\label{ma}
\end{equation}
Here $\vec n$ stands for a certain configuration of $S, I$ and $R$ particles at time $t$.
In a lattice gas description each lattice point is either empty or single occupied leading to 
$n_X(i) = 0,1$ for each type. Using the expansion
\begin{equation}
\mid F(t) \rangle = \sum_{n_i} P(\vec n,t) \mid \vec n \rangle\,.
\label{fo2}
\end{equation} 
 Eq.~(\ref{ma}) can be rewritten as an equivalent 
equation in a Fock-space
\begin{equation}
\partial_t \mid F(t)\rangle = - H \mid F(t) \rangle,
\label{fo1}
\end{equation}
where the operator $H$ is determined in such a manner that its matrix elements correspond to 
those of $\mathcal{L}$. The formal solution of that equation is
\bel{formalsolution}
  \mid F(t) \rangle = \e^{-Ht}  \mid F(0) \rangle.
\ee
This expression gives the probability distribution at time $t$ in terms of the
initial distribution at time $t=0$.

It should be emphasized that the procedure is up to now 
independent of the realization of the basic vectors $\mid \vec n \rangle$. As shown by Doi \cite{doi} the 
average of an arbitrary physical quantity $\mathcal{R}(\vec n)$ can be calculated by the average of the 
corresponding diagonal operator $R(t)$
\begin{equation}
\langle R(t) \rangle = \sum_{n_i} P(\vec n,t) \mathcal{R}(\vec n) = 
\langle s \mid R \mid F(t) \rangle 
\label{fo3}
\end{equation} 
with the summation vector $\langle s \mid = \sum \langle \vec n \mid$. 
The evolution equation for an operator 
$R(t)$ can be cast in a commutator relation which reads
\begin{equation}
\partial_t \langle R \rangle = \langle s \mid [R(t),H]_{-} \mid F(t) \rangle.
\label{kin}
\end{equation}
As the result of the procedure, all the dynamical equations governing the 
classical problem are determined by the structure of the evolution operator 
$H$ and the commutation rules of the operators.\\ 
The evolution operator for the process defined by Eq.~\eqref{mod} reads 
\begin{eqnarray}
- H &=&\beta  \sum_i \left[ b^{\dagger}_{i+1} a_{i+1} + 
b^{\dagger}_{i-1} a_{i-1}  - A_{i+1}(1 - B_{i+1}) - A_{i-1}(1 - B_{i-1}) \right ] B_i\nonumber\\
&+& \gamma \sum_i (b_i - B_i)\,. 
\label{evo1}
\end{eqnarray}
Here $a_i, a^{\dagger}_i$ and $b_i, b^{\dagger}_i$ are the annihilation and creation operators for 
$S$ and $I$ types. The operators $A_i = a^{\dagger}_i a_i $ and $B_i = b^{\dagger}_i b_i$ 
represent the particle number operators with the eigenvalues 
$0$ and $1$. The particle number operator $A_i$ corresponds to the 
occupation variable $n_S(i)$ and $B_i$ corresponds to $n_I(i)$. 

The meaning of the evolution 
operator defined in Eq.~\eqref{evo1} 
is now transparent: The first term on the right hand side describes the annihilation of a susceptible at 
site $i+1$ and a simultaneous creation of an infectious at the same site provided the neighboring lattice 
site $i$ is occupied by an infectious indicated by the number operator $B_i = b^{\dagger}_i b_i$. 
Mathematically this property is manifest in the commutator relation
\begin{equation}
[b_i, b^{\dagger}_j] =  (1 - 2 b^{\dagger}_i b_i ) \delta_{ij}\,.
\label{com}
\end{equation}
and similar rules for the $a$ and $a^{\dagger}$. The operators commute at different lattice sites and 
the anticommute at the same lattice site. The anticommutator rule implies the exclusion principle, i.e. 
the eigenvalues of the particle operators are restricted to $0, 1$ and therefore
the corresponding averages fulfills $0 \leq \langle A_i \rangle \leq 1$. Similar relations hold for
$B_i$. From the definition follows
\bea
\label{expectationS}
\exval{n_S(i,t)} & = & \exval{A_i} \\
\label{expectationI}
\exval{n_I(i,t)} & = & \exval{B_i} 
\eea
and correspondingly
\bel{expectationR}
\exval{n_R(i,t)} = 1-  \exval{A_i} - \exval{B_i}
\ee
for the probability of finding node $i$ in the recovered state. Notice that these
expectation values imply a double average over the initial distribution and over
realizations of the stochastic dynamics.

\subsection{Cluster functions}

Using Eq.~\eqref{kin} and Eq.~\eqref{com} we get
\begin{eqnarray}
\frac{\partial }{\partial t} \langle A_r \rangle &=& -\, \beta \left( \langle A_r B_{r-1} \rangle 
+ \langle A_r B_{r+1} \rangle \right) \nonumber\\
\frac{\partial }{\partial t} \langle B_r \rangle &=&\quad \beta  \left( \langle A_r B_{r-1} \rangle 
+ \langle A_r B_{r+1} \rangle  \right) - \gamma \langle B_r \rangle
\label{evo2}
\end{eqnarray}

These equations involve second-order correlators which hints at the non-linear nature of the problem.
To analyze the situation let us further study the higher order correlators 
appearing in Eqs.~\eqref{evo2}. For illustration we present the result for the two-point correlator
\begin{equation}
\frac{\partial }{\partial t} \langle A_r B_{r+1} \rangle = 
-(\gamma +\beta) \langle A_r B_{r+1} \rangle + 
\beta \left( \langle A_r A_{r+1} B_{r+2} \rangle - \langle B_{r-1} A_r B_{r+1} \rangle \right)
\label{evo3}
\end{equation}
To make a more systematic approach let us define the $n$-point cluster functions 
\begin{eqnarray}
H_r(n) &=& \langle A_r A_{r+1} \dots A_{r+n-1} B_{r+n}\rangle \nonumber\\
G_r(n) &=& \langle B_{r-1}A_r A_{r+1} \dots A_{r + n-1} B_{r+n}\rangle.
\label{evo4}
\end{eqnarray}
Obviously these functions are zero if one of the sites inside of the cluster is
recovered
with probability 1. Furthermore 
the functions $H(n)$ and $G(n)$ are sensitive to the fact that a cluster of 
susceptible individuals is diminished by infection at the border of the cluster. 
The introduction of these cluster functions is the decisive trick of our treatment which
makes the nonlinear problem solvable. 

The cluster equations simplify under the natural 
assumption of a translation invariant initial 
distribution. Then one can drop the $r$-dependence and
after a straightforward calculation only taking into account the 
algebraic properties Eq.~\eqref{com} we end up with the following set of coupled equations for the 
cluster functions 
\begin{eqnarray}
\frac{\partial}{\partial t} H(n) &=& - (\gamma +\beta ) H(n) +\beta 
(H(n+1) - G(n) )\nonumber\\
\frac{\partial}{\partial t} G(n) &=& - 2( \gamma + \beta ) G(n) + 2 \beta G(n+1)\quad {\rm for}\quad n \geq 1 .
\label{high}
\end{eqnarray}
For non-translation invariant distributions the evolution equations for the cluster functions
still close, but retain an extra $r$-dependence.
The meaning of the evolution equations is immediately visible. The cluster described by $H(n)$ decreases by 
infecting node $r$ with rate $\beta$ and recovery of node $r+n$ with rate $\gamma$. Furthermore the 
$H(n)$ cluster grows by increasing its length from $n$ to $n+1$ through infection of node $n+r$. 
The exact evolution equation of the $G$-cluster can be made plausible in similar terms.
This competition between growth and reduction processes of the clusters yields a non-trivial
steady state of the process discussed below.

The second cluster equation can be solved recursively by treating the term $G(n+1)$
as an inhomogeneity of the remaining homogeneous first-order linear ordinary differential
equation. One obtains
\begin{equation}
G(n, t) = \e^{- 2 ( \gamma + \beta)t}  \sum_{l=0}^{\infty} \frac{(2\beta t)^l}{l!} G(l+n, 0)\,,
\label{sol1}
\end{equation}
where $G(n+l,0)$ is an arbitrary initial condition. 

Inserting that solution in the first equation we find in a similar fashion
the solution 
\begin{eqnarray}
H(n,t) &=& \e^{-(\gamma + \beta) t} \sum_{l=0}^{\infty} \frac{(\beta\,t)^l}{l!} H(l+n,0) \nonumber\\
&-&\beta \int_0^t dt' \e^{-(\gamma + \beta) (t - t')} \sum_{l=0}^{\infty} 
\frac{[\beta (t - t')]^l}{l!} G(n+l, t') 
\label{sol4}
\end{eqnarray}

These results are exact for arbitrary translation invariant initial distributions.
For uncorrelated random initial distribution of each class of individuals
the initial conditions for the cluster functions read
\begin{equation}
G(n,0) = n_S(0)^n n_I(0)^2,\quad H(n,0) = n_S(0)^n n_I(0).
\label{in}
\end{equation}
Here the initial densities of infectious and susceptible individuals are
$n_I(0)$ and $n_S(0)$.
Inserting this in Eq.~\eqref{sol1} and Eq.~\eqref{sol4}
results in
\begin{eqnarray}
\label{clusterG}
G(n,t) &=& n_S(0)^n n_I(0)^2 \exp(-\frac{2 t}{\tau}\,)\\
\label{clusterH}
H(n,t) &=& n_S(0)^n n_I(0) \exp(-\frac{t}{\tau}\,)\left[ 1 - \beta n_I(0)\tau
( 1 - \exp(-t/\tau ) \right],
\end{eqnarray}
where the relaxation time $\tau$ is defined by  
\bel{relaxationtime}
\tau = \frac{1}{\gamma +\beta ( 1 - n_S(0))}.  
\ee
We draw attention to the fact that the relaxation time depends on the 
initial conditions through the initial density of susceptibles. This is a consequence
of the highly non-ergodic and far-from-equilibrium nature of the process.
Both cluster functions decay monotonically in time to zero.

\section{Exact solution} 

Using the exact result for the cluster expectation values for random initial
conditions obtained in the previous
section from the quantum approach, we are now in a position to obtain the
exact time evolution for the expected number of individuals of each class. 
In terms of the cluster functions $G(n)$ and $H(n)$ the evolution equations for the quantities 
\be
n_S(t) = \langle A_r \rangle \quad  {\rm and} \quad n_I(t) = \langle B_r \rangle
\ee
read
\begin{eqnarray}
\frac{\partial }{\partial t} n_S(t) &=& - 2 \beta H(1,t)\nonumber\\
\frac{\partial }{\partial t} n_I(t) &=& 2 \beta H(1,t) - \gamma n_I(t).
\label{sol5}
\end{eqnarray}
Notice that $H(1,t)$ is strictly positive for all finite times. Hence the density of
susceptible individuals is strictly monotonically decreasing which follows from the 
fact that no susceptibles are generated
in the process.

The averaged number of susceptible person 
at time $t$ follows immediately from Eqs. \eqref{clusterG}, leading to 
\begin{eqnarray}
n_S(t) = n_S(0) &-& 2\beta \tau n_S(0) n_I(0) \left\{\left[ (1 -  \exp(-\frac{t}{\tau})\right]
\left[ 1 - \beta \tau n_I(0) \right]\right. \nonumber\\
&+& \left.\left[1- \exp(-\frac{2 t}{\tau})  \right] \frac{\beta \tau n_I(0)}{2} \right\}  .
\label{solA}
\end{eqnarray}
The decay of the susceptible persons $S$ is not purely Arrhenius-like but 
given by a superposition of two 
relaxation times. 

In the long time limit we find a nonzero stationary solution 
\be
n_S^\ast \equiv n_S(t \to \infty) = n_S(0) \left[ 1 - \beta \tau n_I(0) \right]^2
\ee
which can be written in the more transparent form
\begin{equation}
\frac{n_S^\ast }{n_S(0)} = \left[ \frac{ \gamma + \beta (1 - n_S(0) - n_I(0)}{ \gamma + \beta (1 - n_S(0))}\right]^2 
\label{stat}
\end{equation}
which makes the dependence on initial conditions and the recovery/infection
ratio $\gamma/\beta$ fully explicit.

In the same manner we find the expected density of infectious persons
\be
\label{solB}
n_I(t) = n_I(0) [ \exp(-\gamma t)  + 2 \beta \tau n_S(0) f(t) ]
\ee
with
\be
f(t) = \frac{\exp(-t/\tau) - \exp(-\gamma t)}{\gamma \tau - 1} \left[ 1 - \beta n_I(0) \tau \right ] 
+ \frac{\beta n_I(0) \tau}{\gamma \tau - 2} \left[ \exp(-2t/\tau) - \exp(-\gamma t)  \right]
\ee
Due to Eq.~\eqref{con} $1 - n_S - n_I$ is the expected density of recovered individuals.

For a comparison with the predictions of the original SIR model defined by the set of differential 
equations \eqref{mfa} the mean-field solution we highlight some features of this mean field model.
From the first equation in Eq.~\eqref{mfa} we conclude that $S(t)$ is a monotonically decreasing function.
Moreover, the last equation implies that the stationary value for the 
infectious class is $I^{*} = 0$. Both properties are shared by our stochastic SIR model. 
In the condition for the existence of a maximum in the number of
infectious individual, the situation is more subtle. Writing the second equation  in Eq.~\eqref{mfa}
in the form
\be
\dot{I} = I (\beta S - \gamma)
\ee
one realizes that a maximum occurs if $S(0) > \gamma/\beta$. It is reached at a time $\Theta$
where $S(\Theta) = \gamma/\beta$. Inserting this into the first equation one may write in
terms of the normalized population densities
$\dot{n}_S(\Theta) = \gamma n_I(\Theta)$. Interestingly, the exact relation \eqref{sol5} asserts 
that in our stochastic model
the maximum occurs at a time determined by the same relation
in the
case of random initial conditions. Hence, our model reproduces several key features of the
original SIR model. For these features, the low connectivity and stochasticity 
are unimportant for a comparison of average behavior of the stochastic dynamics
with the deterministic
behavior of the mean field model. Notice though, that the actual value of
$\Theta$ is not the same in the two models. It is also interesting to observe that
the mean field expression for $\Theta$ can be written in terms of the concentration in the
form $n_S(\Theta) = \gamma/(\beta N)$, i.e. the maximum in $I$ occurs at a time
where an initial concentration of susceptibles of order 1 has almost disappeared
and only a finite total number (of order 1) of susceptibles are left in the population.
This is in contrast to our stochastic model where the maximum in $I$ occurs at a concentration
of susceptibles which is of order 1. In this respect, the mean field model fails to
capture the effects of low connectivity.

For a more detailed analysis of the mean field SIR model
we introduce $g(t) = \ln S(t)$ which due to the first equation in 
Eq.~\eqref{mfa} satisfies $\dot{g} = -\beta I$. Differentiating again and using the second
equation gives an integrable second order equation for $g$.  After one integration one obtains
\be
\frac{d g}{dt} = \beta \, \e^g - \gamma g + \Delta
\ee
where $\Delta$ is an integration constant. In case of the initial conditions $S(0) = S_0, I(0) =I_0 $ and 
$R(0) = 0$ it results $\Delta = \gamma \ln S_0 - \beta (S_0 + I_0)$. Combining the last relation for 
$g(t)$ with Eq.~\eqref{mfa} we then find the relation
\bel{IandS}
I(t) = - \beta S(t) + \frac{\gamma}{\beta} \ln \frac{S(t)}{S_0} + N. 
\ee

In the same manner we find 
\begin{equation}
R(t) = - \frac{\gamma}{\beta} \ln \frac{S(t)}{S_0}\,,
\label{mfa2}
\end{equation}
where $R(t)$ obeys
\be
\frac{dR}{dt} = - \gamma S_0 \exp[-\frac{\beta}{\gamma} R] + \gamma (N - R).
\ee
It corresponds to an overdamped motion 
in a potential
\begin{equation}
\frac{dR}{dt} = - \frac{d U(R)}{dR},\quad {\rm with}\quad U(R) = - \frac{S_0 \gamma^2}{\beta} 
\exp[-\frac{\beta}{\gamma} R] + \frac{\gamma R}{2} (R - 2 N) .
\label{mfa3}
\end{equation}
This equation of motion does not allow for a closed solution in terms of elementary functions.

In the limit $t \to \infty$ Eq. \eqref{IandS} gives a transcendental equation for the
stationary population of 
susceptibles
\be
\beta S^* - \ln \frac{S^*}{S_0} - \beta N = 0.
\ee
This has no solution in closed form, but for large $N$ one obtains
\begin{equation}
S^* \simeq S_0 \exp (-\frac{\beta}{\gamma} N )
\label{mfa1}
\end{equation}
which decays exponentially in  the population size $N$.
This result is strongly different from the exact solution \eqref{stat} where one finds a finite 
stationary value of order 1 even for infinite $N$. 

As a final remark we point out that 
in the mean field approximation one decomposes higher order correlators according to 
$\langle A B \rangle = \langle A \rangle \langle B \rangle $. 
Identifying $\langle A \rangle$ with the 
density of susceptibles $n_S(t)$ and correspondingly $\langle B \rangle = n_I(t)$ we get  
from \eqref{sol5}
mean field equations of the form \eqref{mfa}, but with an infection rate $\beta_{mf} = \beta/N$.
Hence the mean field approximation of our stochastic model yields a deterministic SIR dynamics with
renormalized infection rate $\beta_{mf}$.

\section{Monte Carlo simulation data}

Our exact results are obtained for the thermodynamic limit of infinite population size, and
they are results for a statistical ensemble of processes, averaged both over random initial states and histories.
Here we present Monte Carlo simulation results for single runs of the process
which demonstrate that even if the moderate population size is moderate, fluctuations around the
computed expectation value are rather small. This mean that the computed expectation values
represent the typical behavior that one expects in a single outbreak of the disease.
Only for very small populations the fluctuations around the expected mean become significant. 

We have performed the numerical simulation of the problem 
as follows.  Initially, each site is occupied independently and randomly by a susceptible 
with probability $n_S(0)$ and by an infectious individual with probability
$n_I(0) = 1-n_S(0)$.  For the dynamics we have chosen a random sequential update
algorithm as follows. An arbitrary lattice site $j$ is chosen randomly.
If this site is occupied with an infectious $I$, then the $I$-state decays to $R$ with probability 
$\gamma/(2\beta + \gamma)$. If it does not decay, 
then with equal probability 1/2 an adjacent site on the left or right hand site is chosen. 
If the chosen neighboring site is occupied by 
a susceptible $S$, then $S$ is converted into $I$ with probability 
$2\beta( 2\beta + \gamma\,)$. If
lattice site $j$ is occupied by a susceptible $S$ or the site is recovered nothing happens.
Then a new site is selected randomly and the procedure is repeated.
$N$ such update steps then define one Monte-Carlo time step.
We remark that for an efficient implementation of the process one may keep a
list of coordinates of infectious sites and select sites only from this list. However,
for population sizes of the order of $10^3$ such optimization is irrelevant
for the study of single realizations of the process.

\begin{figure}
\begin{subfigure}[\,$\beta = 0.7$, $\gamma = 0.8$, $n_S(0)= 0.4, n_I(0) = 0.6$\,]
{\includegraphics[width=0.45\linewidth]{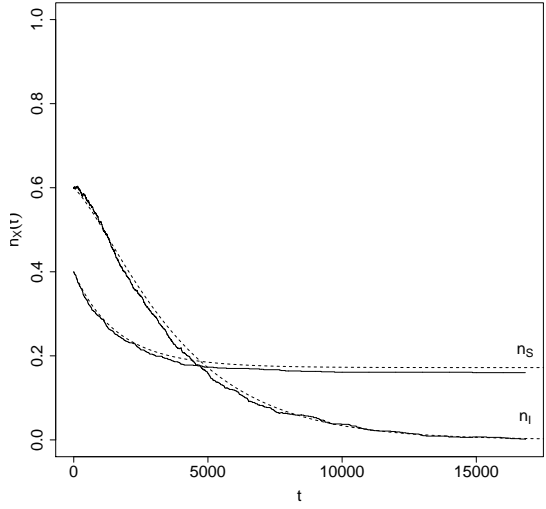}}
\end{subfigure}
\hfill
\begin{subfigure}[\,$\beta = 0.9$, $\gamma = 0.1$, $n_S(0)= 0.8, n_I(0) = 0.2$\,]
{\includegraphics[width=0.45\linewidth]{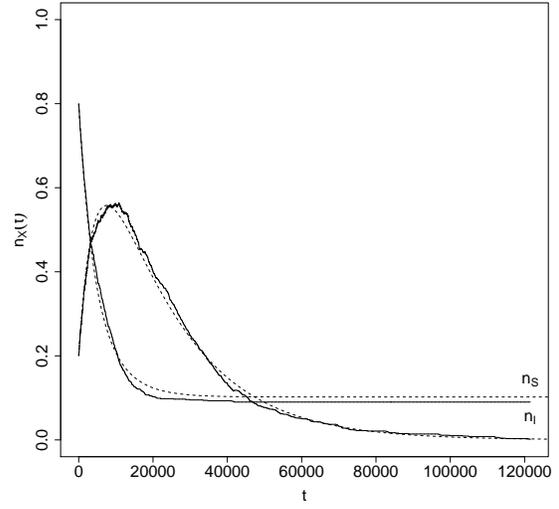}}
\end{subfigure}
\caption{Time evolution of susceptibles and infectious individuals for different rates 
$\beta$ and $\gamma$. The full line corresponds to the simulations, the dashed line represents 
the exact solution. The population size is in both cases $N=10^3$}
\label{fig1}
\end{figure}

In Fig.~\ref{fig1} we show simulation data for two different 
runs with population size 1000, demonstrating the absence (Fig.~\ref{fig1}(a)) 
or presence (Fig.~\ref{fig1}(b)) respectively 
of a maximum in the number of infectious particles. The maximum occurs at 
values of $n_s$ which is of order 1, rather than $1/N$ as predicted by mean field theory. 
The finite limiting value of the susceptible population density is also clearly seen. 
The corresponding mean field value $\approx 400 \exp(-7000)$ would be nearly zero.
For a comparison of this single run
with the computed mean values the corresponding exact expressions
$n_S(t)$ \eqref{solA} and $n_I(t)$ are shown as well. The deviations
are at most in the range of a few percent.
\eqref{solB}. 

\begin{figure}
\begin{subfigure}[\,$N = 10000$]
{\includegraphics[width=0.45\linewidth]{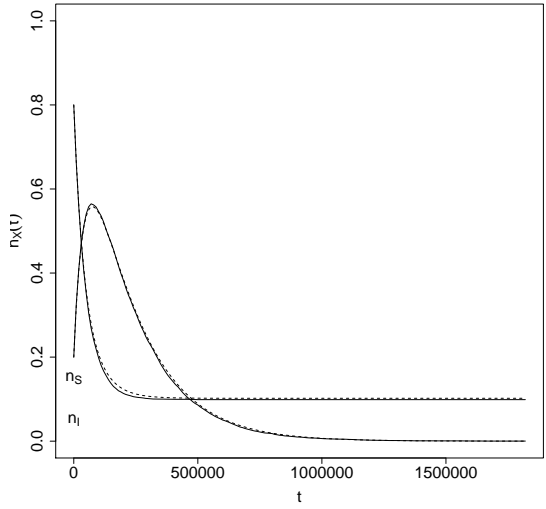}}
\hfill
\end{subfigure}
\begin{subfigure}[\,$N = 100$]
{\includegraphics[width=0.45\linewidth]{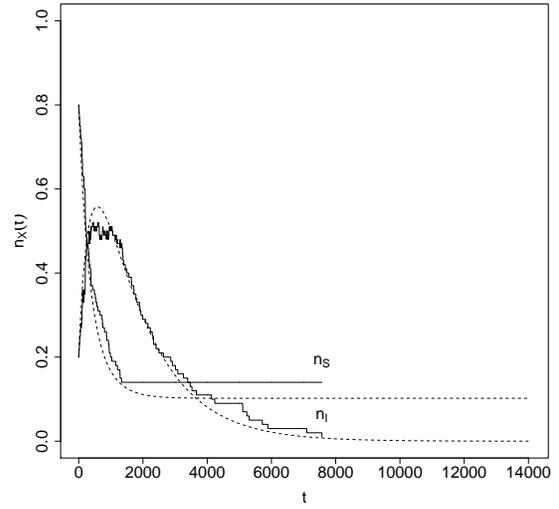}}
\end{subfigure}
\caption{Time evolution of susceptibles and infectious individuals for different population size: 
The rates are in both cases $\beta = 0.9, \gamma = 0.1$, the inital values are $n_S(0) = 0.8, n_I(0) = 0.2$. 
The dashed line is the exact solution.}
\label{fig2}
\end{figure}

In Fig.~\ref{fig2} we show simulation data and the exact 
solution for different total number of individuals. Fig.~\ref{fig2} demonstrates the 
he increasing effect of fluctuations for small population sizes. For population sizes of 
the order of $10^4$ fluctuations become irrelevant.

\section{Conclusions}

\noindent In this work we have analyzed a SIR model for a population
of susceptible $S$, infectious $I$ and recovered $R$ individuals 
evolving under a stochastic dynamics. In order to study the effect of fluctuations
due to incomplete contact between the individuals we have defined the model 
on a linear chain. 
As an appropriate tool we have considered the master equation for
the probability density which we wrote in a quantum formulation in terms of 
second quantized operators. These operators obey the commutation rules of Pauli operators, i.e., 
they commute at different lattice sites and anticommute at the same site.  This property led us to find
a coupled set of evolution equations forcertain cluster functions. These clusters describe the
behavior of 
susceptibles surrounded by infectious individuals at the edges of the clusters and allow
for an exact analytical treatment of
the whole hierarchy of evolution 
equations. We stress that in the exact solution all fluctuations are included. 

Comparing this exact solution with the behavior of the traditional mean field SIR model, 
we find a significant difference. 
Whereas the mean field solution yields a stationary density for the susceptibles 
$n_S^* \simeq \exp (- \beta N/\gamma)$ which depends on the population size and is
extremely small for large $N$, the exact solution reveals a stationary density independent of $N$ 
and of order 1. This shows on a quantitative level how fluctuations due to low connectivity of 
individuals are crucial for understanding the spreading of a disease in the framework of the SIR 
mechanism. 

We remark that by making a mean field approximation to the exact evolution equations
\eqref{sol5} of our model, one obtains a deterministic set of evolution equations similar to those of the
mean-field SIR model, but with an infection rate $\beta_{mf} = \beta/N$. Indeed, inserting
$\beta_{mf}$ in the stationary density of the mean field SIR model, yields a finite stationary
density of susceptibles of order 1, as in our stochastic SIR model. Thus the effect
of low-connectivity model can be qualitatively described by a mean-field model with a
small renormalized infection rate $\beta_{mf}$. Capturing the precise form of the time evolution, however,
is beyond the capabilities of the mean-field description.

The analytical findings are illustrated by numerical simulations which demonstrate that
fluctuations due to finite population size are negligible for population sizes of order
1000 or larger.  We stress that while here we have focused on uncorrelated
initial distributions which are on average spatially homogeneous,
our exact analytical approach can be extended to study the effect of correlations 
and spatial inhomogeneities in the initial distribution. The model remains
exactly solvable also for finite population size.

\begin{acknowledgments}
This work has been supported by the DFG (SFB 418). Two of us (G.M.S and S.T.) are grateful to the 
Weizmann Institute for kind hospitality.  Part of this work was done while G.M.S. was the Weston Visiting
Professor at the Weizmann Institute of Science. G.M.S. also thanks the University of Halle for kind hospitality. 
We thank also Michael Schulz for discussions. 
\end{acknowledgments}


\begin{thebibliography}{90}
\bibitem{f1} The $R$-state may more generally represent an individual
removed permanently from infectious/susceptible process, due to recovery, death or any other mechanism. 
Such alternate interpretations of the $R$-state do not change the mathematical properties of the SIR model
that we study here.
\bibitem{f2} Because of the
close correspondence we use the same
notation $X(t),n_X(t)$ for expectation values in our stochastic model and for the
corresponding quantities in the mean-field SIR
model, but we warn the reader of the difference of meaning in each context.
\end{thebibliography}
\end{document}